\documentclass[%
 superscriptaddress,
preprint,
 amsmath,amssymb,
 aps,
]{revtex4-1}

\usepackage[dvipdfmx]{graphicx}
\usepackage{dcolumn}
\usepackage{bm}

\usepackage{multirow}

\begin{document}

\preprint{APS/123-QED}

\title{Direct observation of the M1 transition between the ground state fine structure splitting of W~VIII}

\author{Momoe Mita}
\affiliation{Institute for Laser Science, The University of Electro-Communications, Chofu, Tokyo 182-8585, JAPAN}

\author{Hiroyuki A. Sakaue}
\affiliation{National Institute for Fusion Science, Toki, Gifu 509-5292, JAPAN}

\author{Daiji Kato}
\affiliation{National Institute for Fusion Science, Toki, Gifu 509-5292, JAPAN}
\affiliation{Department of Fusion Science, SOKENDAI (The Graduate University of Advanced Studies), Toki, Gifu 509-5292, JAPAN}

\author{Izumi Murakami}
\affiliation{National Institute for Fusion Science, Toki, Gifu 509-5292, JAPAN}
\affiliation{Department of Fusion Science, SOKENDAI (The Graduate University of Advanced Studies), Toki, Gifu 509-5292, JAPAN}

\author{Nobuyuki Nakamura}
\affiliation{Institute for Laser Science, The University of Electro-Communications, Chofu, Tokyo 182-8585, JAPAN}

\date{\today}

\begin{abstract}
We present direct observation of the M1 transition between the fine structure splitting in the $4f^{13} 5s^2 5p^6$ $^2F$ ground state of W~VIII.
The spectroscopic data of few-times ionized tungsten ions are important for the future ITER diagnostics, but there is a serious lack of data.
The present study is part of an ongoing effort to solve this lack.
Emission from the tungsten ions produced and trapped in a compact electron beam ion trap is observed with a Czerny-Turner visible spectrometer.
Spectra in the EUV range are also observed at the same time to help the identification of the previously-unreported visible lines.
The observed wavelength $574.47 \pm 0.03$ nm (air), which corresponds to the fine structure splitting of $17402.5 \pm 0.9$ cm$^{-1}$, shows reasonable agreement with the previously reported value $17410 \pm 5$ cm$^{-1}$ obtained indirectly through the analysis of EUV spectra [Ryabtsev {\it et al.}, Atoms {\bf 3} (2015) 273].

\end{abstract}

\maketitle


\section{Introduction}

Tungsten will be used as a plasma-facing material in ITER, and thus is considered to be the main impurity ions in the ITER plasma \cite{Ralchenko2}.
In order to suppress the radiation loss due to the emission of the impurity tungsten ions, it is important to understand the influx and the charge evolution of tungsten ions in the plasma through spectroscopic diagnostics.
The charge states span a wide range from neutral or 1+ near the edge up to neonlike (64+) or higher near the core \cite{Peacock1,Skinner1}.
There is thus a strong demand for spectroscopic data of tungsten ions for a wide range of charge states, and then for a wide range of wavelength.
In particular, it has been recently pointed out that the diagnostics and control of the edge plasma are extremely important for steady state operation of high-temperature plasmas.
Thus the atomic data of relatively low charged tungsten ions are growing significance in the ITER plasma diagnostics \cite{Clementson4}.
However, there is a serious lack of the spectroscopic data for W~VIII - W XXVII as found in the data list compiled by Kramida and Shirai \cite{Kramida1} and also as pointed out by Ralchenko \cite{Ralchenko2}.
Several groups have thus been recently making efforts to solve this lack.
For example, Ryabtsev and co-workers \cite{Ryabtsev1,Ryabtsev2,Ryabtsev3} observed EUV spectra of W~VIII excited in vacuum spark and made detail analysis of the spectra with the aid of a line identification program \cite{Azarov1,Azarov2}.
Through their efforts, it has been confirmed that the ground state configuration of W~VIII is $4f^{13} 5s^2 5p^6$, which had been uncertain in the previous experimental \cite{Veres1} and theoretical \cite{Carlson1,Sugar2} studies due to the competition with $4f^{14} 5s^2 5p^5$.
They also determined the fine structure splitting in the ground state ($^2F_{5/2}$ and $^2F_{7/2}$) to be $17410 \pm 5$~cm$^{-1}$.

These ions attract attention not only from plasma physics but also from the fundamental physics point of view.
Recently, visible transitions in highly charged ions are suggested as potential candidates for a precise atomic clock that can be used to test the time variation of the fundamental physical constant, such as the fine structure constant \cite{Berengut1}.
Some transitions in W~VIII and W~IX are also named as candidates in recent theoretical study by Berengut {\it et al.} \cite{Berengut2}.
However, it has not been identified yet by experiment.

We have been using an electron beam ion trap (EBIT) to observe previously unreported lines of tungsten ions \cite{Komatsu2,Kobayashi2}.
An EBIT produces highly charged ions through successive ionization of trapped ions by a quasi-monoenergetic electron beam. It is thus possible to produce selected charge state ions by tuning the electron beam energy, and thus to obtain simple spectra which are useful to identify previously unreported lines.
In this paper, we present observation of the M1 transition between fine structure splitting in the ground state $4f^{13} 5s^2 5p^6$ $^2F$ of W~VIII.
The result is compared with the previous theoretical and experimental studies.

\section{\label{sec:experiment}Experiment}

The experimental setup and the method of the present measurements were essentially the same as those used in our previous studies \cite{Kobayashi1,Kobayashi2}.
Multiply charged tungsten ions were produced with a compact electron beam ion trap (called CoBIT) at the University of Electro-Communications in Tokyo.
The detailed description on the device is given in the previous paper~\cite{cobit}.
Briefly, it consists essentially of an electron gun, a drift tube, an electron collector, and a high-critical-temperature superconducting magnet.
The drift tube is composed of three successive cylindrical electrodes that act as an ion trap by applying a positive potential (typically 30 V) at both ends with respect to the middle electrode.
The electron beam emitted from the electron gun is accelerated towards the drift tube while it is compressed by the axial magnetic field (typically $\sim 0.08$ T) produced by the magnet surrounding the drift tube.
The compressed high-density electron beam ionizes the ions trapped in the drift tube.
In the present study, tungsten was introduced into the trap through a gas injector as a vapor of W(CO)$_6$.

The M1 transition in W~VIII, which was predicted at 574 nm \cite{Ryabtsev1}, was observed with a commercial Czerny-Turner type of visible spectrometer.
A biconvex lens was used to focus the emission on the entrance slit of the spectrometer.
The diffracted light was detected with a liquid-nitrogen-cooled back-illuminated CCD (Princeton Instruments Spec-10:400B LN).
The wavelength was calibrated using emission lines from several standard lamps placed outside \mbox{CoBIT}.
The uncertainty of the wavelength calibration was estimated to be 0.03~nm including systematics.

Emission in the EUV range was also observed to help the identification of visible lines.
A grazing-incidence flat-field spectrometer \cite{Ohashi4} consisting of a 1200 g/mm concave grating (Hitachi 001-0660) and a Peltier-cooled back-illuminated CCD (Roper PIXIS-XO: 400B) was used.

\section{Results and discussion}

\begin{figure}[t]
\centering
\includegraphics[width=\textwidth]{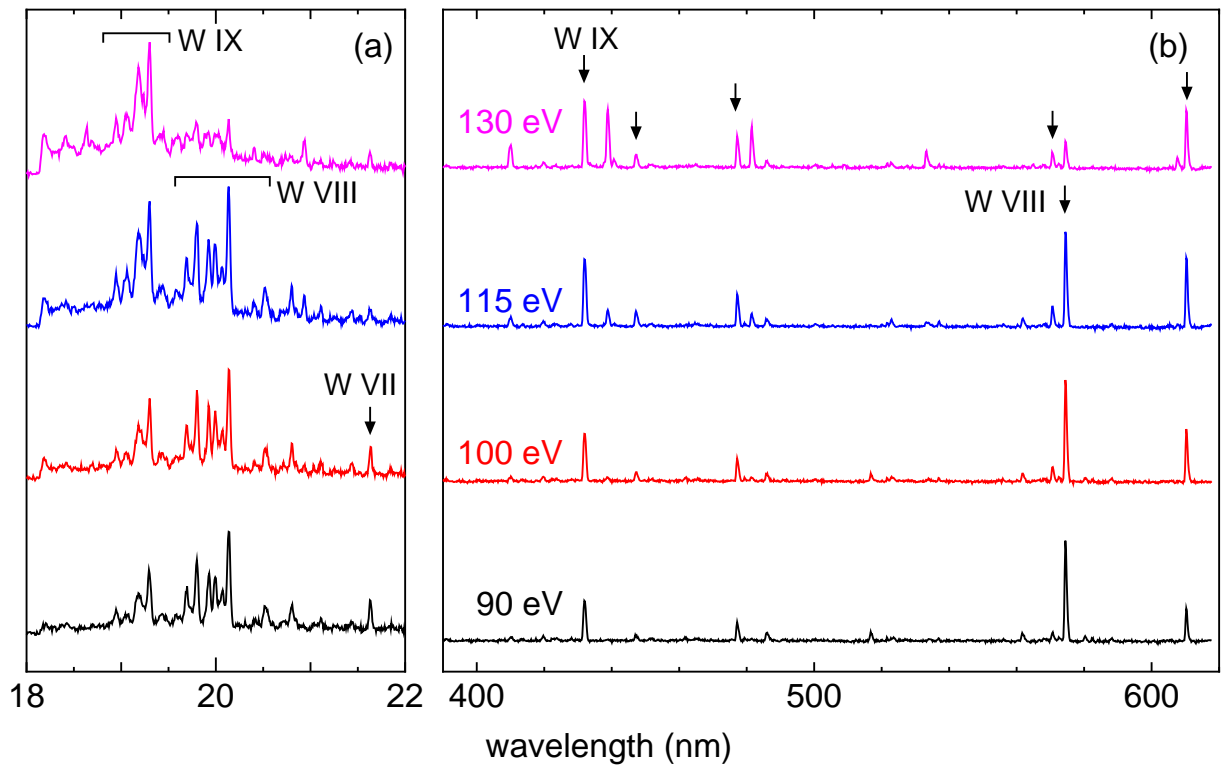}
\caption{\label{fig:spectra}
Spectra of tungsten ions in (a) the EUV range and (b) the visible range, obtained simultaneously at electron energies of 90 (black), 100 (red), 115 (blue), and 130 eV (magenta).}
\end{figure}

Figure \ref{fig:spectra} shows the visible and EUV spectra obtained simultaneously at electron energies of 90, 100, 115, and 130 eV.
Here the electron energy values are simply obtained from the potential difference $V_{\rm cd}$ between the cathode (electron gun) and the middle electrode of the drift tube as $eV_{\rm cd}$.
The actual electron energy (interaction energy between the beam electron and the trapped ions) can be different from the $eV_{\rm cd}$ value due to several reasons, such as the space charge of the electron beam and the trapped ions.
In general, the actual electron energy is lower than the $eV_{\rm cd}$ value by few tens of eV due to the space charge potential of the electron beam; however, it is not sure if it can be applied also in the present study.

In the previous study with the Livermore EBIT, the EUV spectra of tungsten ions were observed with wavelength and electron energy ranges similar to the present study, and the several transitions of W~VII and VIII were identified \cite{Clementson4}.
Our spectra shown in Fig.~\ref{fig:spectra}(a) are generally consistent with their spectra; thus, through the comparison with their spectra the lines of W~VII and VIII can be identified as shown in the figure.
For example, the line at 21.6 nm is assigned as the $5p^6$ - $5p_{1/2}^{-1} 5d_{3/2}$ transition of W~VII.
A cluster of lines near 20 nm mainly corresponds to the $5p$ - $5d$ transitions of W~VIII.
It can be confirmed that the W~VII lines were observed at 90 and 100 eV, and became weak at 115 eV, whereas the W~VIII lines kept increase from 90 to 115 eV and became weak at 130 eV.
This different dependence on electron energy reflects the fact that the charge distribution shifted to higher charge as electron energy increased.
Based on the energy dependence, a cluster of lines near 19nm can be assigned to W~IX although there were not identified in the previous Livermore spectra.
According to the calculation by Ryabtsev {\it et al.} \cite{Ryabtsev1}, a lot of transitions in W~IX are predicted near 18 to 19 nm.
We consider that the cluster of lines near 19 nm in the present EUV spectra corresponds to a part of the predicted lines although the detailed analysis and identification are in progress.
The present identification seems to be inconsistent with the ionization energy values \cite{Kramida1} because W~VIII (122 eV required) and W~IX (141 eV required) lines can be found in the spectrum at an electron energy of 90 eV.
The similar inconsistency is found also in the spectra of the previous Livermore spectra, where they observed W~VII (65 eV required) lines at 43 eV and W~VIII (122 eV required) lines at 72 eV.
Ionization from metastable excited states \cite{Sakoda1} may account for this inconsistency though we are not sure at present.

In the visible spectra, the line observed at 574 nm shows the same electron energy dependence as the W~VIII lines in the EUV spectra.
Thus it can be assigned to the transition between the fine structure splitting of the $^2F$ ground state in W~VIII.
The wavelength is consistent with the energy interval 17410 cm$^{-1}$ previously obtained from the high resolution EUV spectra of vacuum spark \cite{Ryabtsev1}.
The visible spectra shown in Fig.~\ref{fig:spectra} were observed with a low dispersion grating (300 g/mm brazed at 500 nm) to cover a wide visible wavelength range (350 to 620 nm).
We have confirmed that in the observed wavelength range, the line at 574 nm is the only prominent line that can be assigned to W~VIII.
It is reasonable considering the energy level of W~VIII.
According to the calculation by Ryabtsev {\it et al.} (See Fig. 2 of Ref.~\cite{Ryabtsev1}), the two configurations $4f^{14}5p^5$ and $4f^{13}5p^6$ compete for the ground state, and the other configurations have excitation energies of larger than 300,000 cm$^{-1}$, which would make transitions in the EUV range.
The two lowest configurations both have a doublet structure ($^2 P_{1/2, 3/2}$ for $4f^{14}5p^5$ and $^2 F_{5/2, 7/2}$ for $4f^{13}5p^6$), but the energy splitting for $4f^{14} 5p^5 \; {}^2 \! P$ is too large to be observed in the visible range, so that the transition between the doublet levels of $4f^{13} 5p^6 \; {}^2 \! F$ is considered to be the only transition that fall in the visible range.
It may also be possible to have more visible transitions if we can observe transitions between high excited levels.
However, there is less chance to observe such transitions in the low density EBIT plasma.

In order to obtain the transition wavelength, we have also observed the visible spectrum with a higher resolution using a 1200 g/mm grating brazed at 400 nm.
The experimental wavelength obtained from the higher resolution observation is $574.47 \pm 0.03$ nm (air), which corresponds to a fine structure splitting of $17402.5 \pm 0.9$ cm$^{-1}$.
The result is listed in Table~\ref{tab:result} together with previous values.
As seen in the table, our present value shows reasonable agreement with the previous experimental value by Ryabtsev {\it et al.} \cite{Ryabtsev1}.
It is noted that the Ryabtsev value was obtained through the analysis of the EUV spectrum containing a lot of transitions arising from highly excited states to the ground configurations.
On the other hand, the present experiment is the first direct observation of the transition between the fine structure splitting.
The present result has confirmed the reliability of the analysis done by Ryabtsev {\it et al.}

\begin{table}[t]
\caption{\label{tab:result}
Fine structure splitting of the ground state $4f^{13} 5p^6 \; {}^2 \! F$ in W~VIII.
The uncertainty is not given in the theoretical value for Berengut \cite{Berengut1}.
The experimental value by Ryabtsev \cite{Ryabtsev1} was obtained indirectly through the analysis of EUV spectra.
}
\vspace{2mm}
\small 
\centering
\begin{tabular}{cccc}
\hline
& \textbf{year}	& \textbf{exp or th} & \textbf{energy (cm$^{-1}$)} \\
\hline
\hline
Kramida \cite{Kramida1} & 2009			& theory & $17440 \pm 60$\\
Berengut \cite{Berengut1} & 2009			& theory & $18199$\\
Ryabtsev \cite{Ryabtsev1} & 2015			& exp & $17410 \pm 5$\\
Mita (present)		& 2016			& exp & $17402.5 \pm 0.9$\\
\hline
\end{tabular}
\end{table}

In summary, we have observed the visible M1 transition between the ground state fine structure splitting of W~VIII.
The energy splitting obtained from the transition wavelength has shown good agreement with the value obtained in the previous study where the energy interval was indirectly obtained from EUV spectra.
We have also found previously unreported lines of W~IX both in the EUV and visible region, which are currently in analysis and will be published elsewhere.


\end{document}